\documentclass[11pt,a4paper]{llncs}
\usepackage{amsmath}
\usepackage{amssymb}
\usepackage{graphicx}
\usepackage{marvosym}
\usepackage{fancyhdr}

\usepackage{lipsum}
\usepackage{flushend}
\usepackage[hyphens]{url}
\usepackage{hyperref}
\usepackage[hyphenbreaks]{breakurl}

\usepackage{geometry}
\usepackage{tabularx}
\newcommand{\keywords}[1]{\par\addvspace\baselineskip
\noindent\keywordname\enspace\ignorespaces#1}

\fancypagestyle{firstpage}{\fancyhf{}
}

\begin{document}

%\mainmatter  % start of an individual contribution

% first the title is needed
\title{\LARGE{``So Am I Dr. Frankenstein? Or Were You a Monster the Whole Time?'': Mitigating Software Project Failure With Loss-Aversion-Aware Development Methodologies}}

% a short form should be given in case it is too long for the running head
%\titlerunning{Lecture Notes in Computer Science: Authors' Instructions}

% the name(s) of the author(s) follow(s) next
%
% NB: Chinese authors should write their first names(s) in front of
% their surnames. This ensures that the names appear correctly in
% the running heads and the author index.
%
\author{\large{Junade Ali}}
\institute{\large{Engprax Ltd,\\ Edinburgh, Scotland, UK}}

%\author{Alfred Hofmann%
%\thanks{Please note that the LNCS Editorial assumes that all authors have used
%the western naming convention, with given names preceding surnames. This determines
%the structure of the names in the running heads and the author index.}%
%\and Ursula Barth\and Ingrid Haas\and Frank Holzwarth\and\\
%Anna Kramer\and Leonie Kunz\and Christine Rei\ss\and\\
%Nicole Sator\and Erika Siebert-Cole\and Peter Stra\ss er}
%
%\authorrunning{Lecture Notes in Computer Science: Authors' Instructions}
% (feature abused for this document to repeat the title also on left hand pages)

% the affiliations are given next; don't give your e-mail address
% unless you accept that it will be published
%\institute{Springer-Verlag, Computer Science Editorial,\\
%Tiergartenstr. 17, 69121 Heidelberg, Germany\\
%\mailsa\\
%\mailsb\\
%\mailsc\\
%\url{http://www.springer.com/lncs}}

%
% NB: a more complex sample for affiliations and the mapping to the
% corresponding authors can be found in the file "llncs.dem"
% (search for the string "\mainmatter" where a contribution starts).
% "llncs.dem" accompanies the document class "llncs.cls".
%

%\toctitle{Lecture Notes in Computer Science}
%\tocauthor{Authors' Instructions}

\maketitle

\thispagestyle{firstpage}

\begin{abstract}
Case studies have shown that software disasters snowball from technical issues to catastrophes through humans covering up problems rather than addressing them and empirical research has found the psychological safety of software engineers to discuss and address problems to be foundational to improving project success. However, the failure to do so can be attributed to psychological factors like loss aversion. We conduct a large-scale study of the experiences of 600 software engineers in the UK and USA on project success experiences. Empirical evaluation finds that approaches like ensuring clear requirements before the start of development, when loss aversion is at its lowest, correlated to 97\% higher project success. The freedom of software engineers to discuss and address problems correlates with 87\% higher success rates. The findings support the development of software development methodologies with a greater focus on human factors in preventing failure.
\keywords{Software development methodologies, Agile software, loss aversion, socio-technical systems.}
\end{abstract}

%\begin{abstract}
%The abstract should summarize the contents of the paper and should
%contain at least 70 and at most 150 words. It should be written using the
%\emph{abstract} environment.
%\keywords{We would like to encourage you to list your keywords within
%the abstract section}
%\end{abstract}

\section{Introduction}
Case studies of catastrophic software projects \cite{killercomputers} has found a common theme is that minor technical problems will be covered-up rather than addressed until they snowball into disaster. This phenomena can be psychologically explained through \textit{loss aversion}, where research has found that humans will feel the pain of a loss greater than the pleasure of a gain \cite{oelrich2019making,blake2021quantifying}.

There has been limited research of the efficacy of software engineering methodologies \cite{ranete2021best} and whilst there has recently been limited research of using emotion-oriented approaches to handling requirements changes in Agile teams \cite{madampe2023framework}, there has no work considering how to mitigate the impact of loss aversion in software engineering teams.

Previous research conducted using different methodologies concurs on the importance of software engineers having the psychological safety to discuss and address problems \cite{48455,darkside}, but organisational transformation initiatives often struggle to succeed \cite{oludapo2024so}.

In this paper we will consider an alternative approach to mitigating the effects of loss aversion in software engineering teams by presenting a loss aversion aware software engineering methodology we refer to as \textit{Impact Engineering}. We empirically evaluate this approach using research consisting of 600 software engineers in the UK and USA.

\section{Related Work}
There has been little empirical evaluation of software engineering methodologies in scientific literature. A 2021 tertiary analysis \cite{ranete2021best} investigating scientific evidence of the Agile methodology found ``the evidence for the Agile methodology is scarce at best.''

Over recent years, some teams have attempted to conduct large-scale research of software engineers to understand on a more empirical basis what techniques work and which don't. \cite{48455} and \cite{offerman2022study} provide examples of where large-scale surveys have been used to attempt to measure the adoption of DevOps techniques against measurement frameworks which prioritise speed.

\cite{48455} used ``snowball sampling'' whereby respondents are asked to share the study on in the aim that a large enough sample will be developed to be representative, these results were used to conduct annual ``State of DevOps'' reports. Another key limitation of \cite{48455} is that the research measures outcomes against metrics which prioritise speed of delivering software over other factors, despite representative opinion polling of British adults finding speed of software delivery to be the least important factor \cite{darkside}, with factors like data security, not having serious bugs and data accuracy being factors most likely to matter ``to a great extent'' to the public.

However, over recent years more sophisticated techniques have developed in conducting representative sampling of various populations for scientific research. For example, \cite{daas2021charis} shows how research conducted by the polling firm IPSOS Mori was used measure adherence to public health measures during the COVID-19 pandemic. Such polling techniques use large-preassembled opinion polling panels with diverse participants who can be reached very quickly rather than needing to assemble bespoke samples, this allows for the reach to those who ordinarily would not self-select to answering responses by snowball methodology panels.

In 2021 the author of this paper also adopted these techniques in software engineering populations by commissioning the research firm Survation to study the impact of COVID-19 on burnout amongst software engineers \cite{ali2021study}.

Notwithstanding the different methodologies, the authors of \cite{48455} and \cite{darkside} concurred that psychological safety (i.e. the ability to discuss and address problems) is fundamental to software engineering research.

In the realm of psychological safety there has been a developing understanding of the role that loss aversion plays in restraining individuals from addressing challenges. \cite{oelrich2019making} conducted a study amongst whistleblowers which found that the intention to whistleblow is more heavily reduced by losses than increased by gains. This concept stems from findings in prospect theory which have found that humans will experience the pain of a loss greater than the pleasure of a gain (i.e. for some it can take the gain of \$2 to make up for the loss of losing \$1).

Whilst \cite{oelrich2019making} is a small-scale study of 39 university students, \cite{blake2021quantifying} commissioned the opinion polling firm YouGov to conduct a large-scale UK study to quantify loss aversion identifying ``average aversion to a loss of \pounds500 relative to a gain of the same amount is 2.41''. The researchers also found loss aversion to be affected by demographic factors such as education, age, income and savings. \cite{darkside} conducted such polling amongst software engineers in the UK finding that 75\% reported facing retaliation for reporting wrongdoing at work, with fear of retaliation from management the leading reason for not reporting wrongdoing.

Our previous work in \cite{killercomputers} looked at case studies of catastrophic technology failures such as the UK Post Office scandal (described as the largest miscarriage of justice in British history), the Toyota unintended acceleration bug (with 537 claims settled for crashes that killed or seriously injured people since 2014) and THERAC-25 (faulty software causing fatal radiation overdoses in hospitals). Our analysis of these case studies is that the reason such disasters occur is that the problems start with a technical engineering issue (typically a poor requirements engineering process before the software development starts), however the cover-up of these issues leads to disaster. This work is broadly compatible with work in the field of resilience engineering which looks to view software systems as \textit{socio-technical} systems where the humans play an essential role in the safety of software systems \cite{said2019new}.

Finally, whilst digital transformations have been known to fail often \cite{oludapo2024so}, there is emerging research that emotional factors play a significant role increasing success rates. Unpublished research conducted jointly by EY and the University of Oxford surveyed 935 senior business leaders and their direct reports, alongside 1,127 workforce members more generally across 23 countries and 16 industry sectors finding that prioritising emotional support during business transformations led $2.6x$ higher success rates \cite{sbs}. \cite{desteno2014gratitude} has found that priming participants to feel emotions of gratitude enhances their ability to delay gratification now for a bigger reward later. In a neutral or happy state, participants required only \$55 on average to forego a reward of \$85 in three months, however when participants were primed to feel gratitude the amount raised to \$63.

In software specifically, \cite{madampe2023framework} has identified the importance of focussing on emotions when handling requirements changes in teams using Agile software development methodologies, however there appears to be no research considering the role of loss aversion in requirements change.

A key gap which exists in literature is understanding whether software development methodologies which focus on mitigating the effects of loss aversion are more likely to produce better commercial outcomes.

\section{Preliminary Research}
Prior to conducting our main study, we first conducted an initial research exercise to measure if the commercial outcomes our study looks to measure are agreeable to business decision makers. Initial market research work we conducted with Haystack found business decision makers were reporting they felt the goal of their software engineering teams were to ``to deliver high-quality software on time'' (not necessarily as quickly as possible though). However we sought to experimentally validate this before using it for our study design.

With J.L. Partners we polled 500 business decision makers in the UK and USA via online panel from the 28\textsuperscript{th} - 29\textsuperscript{th} November 2023. J.L. Partners is a member of the British Polling Council and abides by its rules. Table \ref{tab:employee-count} shows responses by participants to the number of employees work for their firm.

\begin{table}[h]
\centering
\begin{tabular}{|l|l|l|l|}
\hline
        & All  & United Kingdom & United States \\ \hline
0-9     & 5\%  & 6\%            & 5\%           \\ \hline
10-49   & 6\%  & 8\%            & 3\%           \\ \hline
50-99   & 7\%  & 11\%           & 3\%           \\ \hline
100-499 & 25\% & 22\%           & 28\%          \\ \hline
500-999 & 41\% & 40\%           & 42\%          \\ \hline
1000+   & 16\% & 13\%           & 19\%          \\ \hline
\end{tabular}
\caption{Responses to: ``How many employees does your firm employ?''}
\label{tab:employee-count}
\end{table}

Table \ref{tab:agreement-goal} shows agreement amongst business decision makers to the statement: ``The goal of a software engineering team is to deliver high-quality software on time''. We find that 98\% of respondents agreed and 0\% expressed disagreement.

\begin{table}[h]
\centering
\begin{tabular}{|l|l|l|l|}
\hline
                           & All  & UK   & USA  \\ \hline
Strongly agree             & 64\% & 65\% & 62\% \\ \hline
Somewhat agree             & 34\% & 33\% & 34\% \\ \hline
Neither agree nor disagree & 2\%  & 1\%  & 3\%  \\ \hline
Somewhat disagree          & 0\%  & 0\%  & 0\%  \\ \hline
Strongly disagree          & 0\%  & 0\%  & 0\%  \\ \hline
Dont know                  & 0\%  & 0\%  & 0\%  \\ \hline
\end{tabular}
\caption{Responses to: ``To what extent do you agree with the following statement- The goal of a software engineering team is to deliver high-quality software on time''}
\label{tab:agreement-goal}
\end{table}

\section{Methodology}
We proceeded with J.L. Partners to poll 600 software engineers (350 in the USA and 250 in the UK) on their experiences with successful and unsuccessful projects.

Fieldwork was conducted from 3\textsuperscript{rd} - 7\textsuperscript{th} May 2024 by online panel. Respondents were first asked: ``Thinking about the last software project you encountered was it successfully delivered on-time and on-budget, to a high standard of quality?'' Respondents were able to select the following two options:

\begin{itemize}
  \item Yes, the project was delivered successfully
  \item No, the project encountered some challenges
\end{itemize}

481 engineers reported on successful projects, and 119 on failed projects. This sampling strategy allowed a comparative analysis of success rates across engineering practices. These results are shown in Table \ref{tab:software-eng-success-rates}.

\begin{table}[h]
\centering
\begin{tabular}{|l|l|l|l|}
\hline
                    & All  & UK   & USA  \\ \hline
Successful project  & 80\% & 77\% & 82\% \\ \hline
Challenged delivery & 20\% & 23\% & 18\% \\ \hline
\end{tabular}
\caption{Responses to ``Thinking about the last software project you encountered was it successfully delivered on-time and on-budget, to a high standard of quality?''}
\label{tab:software-eng-success-rates}
\end{table}

From here, respondents were then asked to make a binary choice to the following questions about the methodologies used:

\begin{itemize}
  \item Were the project requirements clear before the software development process had begun?
  \item Did the project have a complete specification or requirements document before the development started?
  \item Did you have to make significant changes to the requirements late in the development process?
  \item Were the project requirements accurately based on the real-world problem?
  \item Did you have to work on more than one project at the same time?
  \item Did you feel you were able to discuss and address problems quickly when they emerged during the project?
\end{itemize}

Whilst we analysed the results using individual engineering practices, we also grouped techniques together to be analysed. For our ``Impact Engineering'' (IE) methodology, we looked at cases where the following practices were used: clear requirements before development starts, freedom to discuss and address problems, requirements accurate to real-world problem, a complete specification before development starts and no late requirements changes.

We also consider a methodology called ``Agile Requirements Engineering'' (ARE) where respondents reported development starting before clear requirements, no complete specification and significant changes late in development.

It is worth noting however there is some disagreement as to whether these are truly `Agile' practices. One co-author of the Agile Manifesto has proclaimed that the goal of Agile was to create a process ``which doesn't depend on the requirements being stable'' \cite{agile-retake} whilst another signatory to the Agile Manifesto has proclaimed \cite{holub}: ``Why are people so convinced that waterfall up-front (functional) requirements are essential? They're a disease. They're at the core of waterfall thinking. There is no place for them when you work with agility.'' However, another co-author of the Agile Manifesto has disagreed with these claims \cite{speed}. A further critique of deprecation of requirements in the Agile methodology can be found in \cite{meyer2014agile} and \cite{killercomputers} has found that in the Post Office scandal witnesses at the public inquiry were of the view that there was no specification due to the project being conducted using an Agile methodology, namely Rapid Application Development (albeit some would claim this implementation is therefore faulty).

We finally also consider a third group solely of whether respondents reported working on one project at a time versus multiple (i.e. Lean development, given a key tenant of the lean methodology is reducing work-in-progress).

\section{Personal Safety Risks}
Ironically, working on this research we were exposed to loss aversion in those who advocate methodologies which seem to dependent on low levels of loss aversion to succeed.

Since commencing this research, those involved in this research have faced harassment and hacking attempts. In three instances this warranted legal investigation and all but one of those perpetrating the conduct were found to be authors of books advocating Agile, DevOps and Digital Transformation methodologies with the other individual being an Agile practitioner. Of the six individuals identified three were in the United States, one in Canada and two in England. All were male. Most had connections to the leading publisher of DevOps books.

As our organisation engages in security sensitive work day-to-day, we already had suitable operational security practices in place and these threats no longer appear to be active. Nevertheless, we feel it's useful to add this note such that academic researchers can take steps to protect themselves from such conduct both psychologically and physically.

This situation is of course incredibly ironic: as a methodology, Agile, DevOps and Digital Transformation methodologies require low levels of psychological loss aversion to succeed in reality. However, in conducting this research we found those who advocate such methodologies the loudest appeared to fall victim to the phenomenon themselves.

This underscores a point made in \cite{festinger} on the effect of \textit{cognitive dissonance} psychologically: ``Tell him you disagree and he turns away. Show him facts or figures and he questions your sources. Appeal to logic and he fails to see your point."

Whilst this paper does not explore the underlying psychology of this phenomena, it may be worth future research considering whether amongst populations of who advocate such methodologies there is a heightened prevalence of \textit{malignant narcissism} given the manipulativeness, callousness and hostility to criticism seen and how the stories of transformation may appeal to self-glorifying delusions of grandeur \cite{day2022pathological,skoler1998archetypes}.

\section{Results}

The key results from our study is that across our population software engineers reported that projects which had clear requirements before they began were 97\% more likely to succeed, the single biggest factor to improving success. This was followed by the ability to discuss and address problems having an 87\% improvement in success. The other results can be seen in Table \ref{tab:success-increase-practices}.

\begin{table}[h]
\small
\centering
\begin{tabularx}{\columnwidth}{|X|X|}
\hline
Practice & Increase in Success (\%) \\ \hline
Clear requirements at start & 97\% \\ \hline
Psychological safety & 87\% \\ \hline
Real-world aligned requirements & 54\% \\ \hline
Complete spec at start & 50\% \\ \hline
No late-stage changes & 7\% \\ \hline
Single project focus & -3\% \\ \hline
\end{tabularx}
\caption{Increase in success by engineering practice}
\label{tab:success-increase-practices}
\end{table}

Interestingly, we found no statistically significant difference between project success in those claiming to work on one project at a time versus multiple ($p = 0.56$). We do find however that projects adopting the five practices we refer to as ``Impact Engineering'' had a 54\% lower failure rate than the baseline. Table \ref{tab:failure-rates-methodology} shows the failure rate for each methodology studied (Agile Requirements Engineering, Lean and Impact Engineering) alongside the increase in failure rate.

\begin{table}[h]
\small
\centering
\begin{tabularx}{\columnwidth}{|X|X|X|X|X|}
\hline
Methodology & Fail Rate (\%) & Increase (\%) & T-Stat & P-Value \\ \hline
ARE & 65\% & 268\% & 4.94 & 3.83E-5 \\ \hline
Lean & 21\% & 11\% & 0.59 & 0.56 \\ \hline
IE & 10\% & -64\% & -4.15 & 4.11E-5 \\ \hline
\end{tabularx}
\caption{Failure rates by methodology}
\label{tab:failure-rates-methodology}
\end{table}

Finally, Table \ref{tab:practice-prevalence-uk-usa} shows the prevalence in engineering practices between the UK and USA. We find that the biggest difference in engineering practice between the countries is that software engineers in the UK were 13\% less likely to feel they were able to discuss and address problems than those in the US.

\begin{table}[h]
\small
\centering
\begin{tabularx}{\columnwidth}{|X|X|X|}
\hline
Practice & UK (\%) & USA  (\%) \\ \hline
Psychological safety & 79\% & 90\% \\ \hline
Single project focus & 61\% & 71\% \\ \hline
Late requirement changes & 65\% & 59\% \\ \hline
Complete spec at start & 88\% & 83\% \\ \hline
Clear requirements at start & 92\% & 88\% \\ \hline
Real-world aligned requirements & 83\% & 86\% \\ \hline
\end{tabularx}
\caption{Prevalence of practices in the UK and USA}
\label{tab:practice-prevalence-uk-usa}
\end{table}

\section{Limitations \& Future Work}
Our research here has collected empirical data of reported project success rates based on a hypothesis we have observed on the reasons for catastrophic software failure. In future, it may be helpful to conduct a randomised control trial of project success rates as different practices are applied. This will help ensure any differentiation between correlation and causation is better controlled for. Furthermore, future studies may look to measure success rates from the perspective of stakeholders or using other metrics.

Inevitably, such studies may not be able to garner the same population size as we have been able to do here using opinion polling techniques, which have seen vast progress in recent years, but such smaller scale studies may nevertheless provide useful evidence to help control other factors than those already addressed for this study.

Our unique contribution in this work is a psychologically-aware insight into software project failure. Other research in the future may well endeavour to investigate which other psychological factors play a role in software project failure. By looking beyond technical factors, there is a vast sea of opportunity to study and investigate.

\section{Analysis \& Conclusion}
In this study we have looked to understand the role that various factors, including a lack of upfront requirements, plays in software project failure.

Case studies have shown that software disasters can snowball into catastrophes when problems are not addressed and are instead covered-up \cite{killercomputers}. Psychological factors, a key being loss aversion, can incentivise organisations to go down this path despite the disastrous consequences \cite{oelrich2019making}.

Yet, software methodologies have historically disregarded this psychological factor, instead assuming that humans will want to address problems at the earliest opportunity. Previous research of varying methodologies have found the psychological safety to raise and discuss issues to be key to improved software delivery \cite{48455,darkside}. However, it is no doubt optimistic to believe such significant cultural and psychological change can be achieved in all organisations. As we find, even between the UK and USA, psychological safety remains the biggest difference in engineering practice between the two countries.

In this paper, we focus on a different way forward. Instead of attempting to achieve a cultural change that is at best challenging, we instead focus on a more pragmatic approach: limit the triggers of loss aversion from the outset through robust requirements and then aim for psychological safety where it is needed. In our research, the sole factor we found that would increase the success rate of a project more so than psychological safety was having clear requirements from the outset.

Against our hypothesis, the results that merely having clear requirements, even when they change late in development or are not aligned with the real-world, seems to have a significant role in software project success. This suggests that having a gate before the start of a project to discuss and address problems, when loss aversion is at its lowest, allows for the greatest improvement in success rates.

The song Dr. Frankenstein by RedHook \cite{redhook} contains the refrain: \textit{``So am I Dr. Frankenstein? Or were you a monster the whole time?''} For catastrophic software projects, this paper contends that software projects become disasters when humans fail to address smaller scale technical problems. Existing software engineering methodologies fail to address that humans are not machines and psychological factors impede the ability to address problems. Having identified this anachronism in existing methodologies, this paper presents how we can act upon it by using requirements before the start of a project to provide the opportunity to address problems when loss aversion is at its lowest. 

\bibliographystyle{plain}
\bibliography{bls_final}

\begin{thebibliography}{10}

\bibitem{sbs}
``prioritising emotions is the key to success for business transformation, groundbreaking research finds", 2022.
\newblock Available from \url{https://www.sbs.ox.ac.uk/news/prioritising-emotions-key-success-business-transformation-groundbreaking-research-finds}.

\bibitem{ali2021study}
Junade Ali.
\newblock Study to understand the impact of covid-19 on software engineers, 2021.
\newblock Available from \url{https://haystack-books.s3.amazonaws.com/Study+to+understand+the+impact+of+COVID-19+on+Software+Engineers+-+Full+Report.pdf}.

\bibitem{darkside}
Junade Ali.
\newblock The dark side of software development, 2023.
\newblock Available from \url{https://engprax.s3.eu-west-2.amazonaws.com/The+Dark+Side+of+Software+Development.pdf}.

\bibitem{killercomputers}
Junade Ali.
\newblock {\em How to Protect Yourself from Killer Computers: From the Post Office Scandal to Artificial Intelligence}.
\newblock Engprax Ltd, 2024.

\bibitem{blake2021quantifying}
David Blake, Edmund Cannon, and Douglas Wright.
\newblock Quantifying loss aversion: evidence from a uk population survey.
\newblock {\em Journal of Risk and Uncertainty}, 63(1):27--57, 2021.

\bibitem{daas2021charis}
Chris~D. Daas, Gill Hubbard, Marie Johnston, and Duncan Dixon.
\newblock Protocol of the covid-19 health and adherence research in scotland (charis) study: understanding changes in adherence to transmission-reducing behaviours, mental and general health, in repeated cross-sectional representative survey of the scottish population.
\newblock {\em BMJ Open}, 11(2):e044135, 2021.

\bibitem{day2022pathological}
Nicholas~JS Day, Michelle~L Townsend, and Brin~FS Grenyer.
\newblock Pathological narcissism: An analysis of interpersonal dysfunction within intimate relationships.
\newblock {\em Personality and Mental Health}, 16(3):204--216, 2022.

\bibitem{desteno2014gratitude}
David DeSteno, Ye~Li, Leah Dickens, and Jennifer~S Lerner.
\newblock Gratitude: A tool for reducing economic impatience.
\newblock {\em Psychological science}, 25(6):1262--1267, 2014.

\bibitem{festinger}
Leon Festinger, Henry~W. Riecken, and Stanley Schachter.
\newblock When prophecy fails.
\newblock 1956.
\newblock \url{https://doi.org/10.1037/10030-000}.

\bibitem{48455}
Nicole Forsgren, Dustin Smith, Jez Humble, and Jessie Frazelle.
\newblock 2019 accelerate state of devops report.
\newblock Technical report, 2019.
\newblock Available from \url{http://cloud.google.com/devops/state-of-devops/}.

\bibitem{agile-retake}
Martin Fowler, Katherine Kirk, Jez Humble, Dave Thomas, and Tatiana Badiceanu.
\newblock A retake on the agile manifesto.
\newblock In {\em GOTO Conference Aarhus}. Trifork, 2014.
\newblock Available from \url{https://www.youtube.com/watch?v=zNvmjPzdqKc}.

\bibitem{holub}
Allen Holub.
\newblock ``why are people so convinced that waterfall up-front (functional) requirements are essential? they're a disease. they're at the core of waterfall thinking. there is no place for them when you work with agility.", 2024.
\newblock Available from \url{https://x.com/allenholub/status/1821654832996086127}.

\bibitem{redhook}
Emmy Mack, Craig Wilkinson, and Stevie Knight.
\newblock Redhook - dr. frankenstein ft. holding absence.
\newblock Audio, 2024.

\bibitem{madampe2023framework}
Kashumi Madampe, Rashina Hoda, and John Grundy.
\newblock A framework for emotion-oriented requirements change handling in agile software engineering.
\newblock {\em IEEE Transactions on Software Engineering}, 49(5):3325--3343, 2023.

\bibitem{meyer2014agile}
Bertrand Meyer.
\newblock {\em Agile!: The good, the hype and the ugly}.
\newblock Springer Science \& Business Media, 2014.

\bibitem{oelrich2019making}
Sebastian Oelrich.
\newblock Making regulation fit by taking irrationality into account: the case of the whistleblower.
\newblock {\em Business Research}, 12(1):175--207, 2019.

\bibitem{offerman2022study}
Tyron Offerman, Robert Blinde, Christoph~Johann Stettina, and Joost Visser.
\newblock A study of adoption and effects of devops practices.
\newblock In {\em 2022 IEEE 28th International Conference on Engineering, Technology and Innovation (ICE/ITMC) \& 31st International Association For Management of Technology (IAMOT) Joint Conference}, pages 1--9. IEEE, 2022.

\bibitem{oludapo2024so}
Samson Oludapo, Noel Carroll, and Markus Helfert.
\newblock Why do so many digital transformations fail? a bibliometric analysis and future research agenda.
\newblock {\em Journal of Business Research}, 174:114528, 2024.

\bibitem{ranete2021best}
Andrei Ranete.
\newblock ``best practice" without evidence--agile software methodology as example.
\newblock In {\em Norsk IKT-konferanse for forskning og utdanning}, number~4, 2021.

\bibitem{said2019new}
Saloua Said, Hafida Bouloiz, and Maryam Gallab.
\newblock A new structure of sociotechnical system processes using resilience engineering.
\newblock {\em International Journal of Engineering Business Management}, 11:1847979019827151, 2019.

\bibitem{skoler1998archetypes}
Glen Skoler.
\newblock The archetypes and psychodynamics of stalking.
\newblock In {\em The psychology of stalking}, pages 85--112. Elsevier, 1998.

\bibitem{speed}
Richard Speed.
\newblock Study backer: Catastrophic takes on agile overemphasize new features, 2024.
\newblock Available from \url{https://www.theregister.com/2024/08/07/agile_catastrophes_risk_undermining_the/}.

\end{thebibliography}

\vspace{2cm}
\pagebreak

\section*{Acknowledgements}
The author would like to thank J. L. Partners (in particular Tom Lubbock and Julian Gallie) for their support in conducting the fieldwork of the studies described here (and Survation for their assistance with previous fieldwork). The author would like to further thank Haystack (\url{https://usehaystack.io}, in particular Julian Colina and Kan Yilmaz) for their commercial engagement in earlier work conducting market research into the challenges associated with software engineering methodologies.

In memory of Vu Long Tran.

\section*{Author}
\noindent {\bf Junade Ali} is a software engineer and cybersecurity researcher whose technology has been adopted in a variety of areas - from transportation systems to critical internet infrastructure, including being built into products by Apple and Google. He earned his PhD in cryptography after studying for a computer science masters degree aged 17. Aged 23, he earned Chartered Engineer status, the terminal regulatory status for engineers in the UK, and in 2024 aged 27 he became a Fellow of the Institution of Engineering and Technology, the youngest ever Fellow of a professional engineering institution on record.\\

\end{document}